# Quantum resonance scheme to determine the gravitational constant *G*

Zhi Ping Li[1], Xin Li[2]

[1]*1065, Don Mills Road, Toronto, Ontario M3C1X4, Canada*

Electronic address: zhiping_lee@hotmail.com

[2]*College of Information Science and Engineering, Ocean University of China, Qingdao 266100, China*

**Abstract:** Building on the principles of quantum electrodynamics and special relativity, we develop a semi-classical QED theory of gravitation. The experiment model for theoretical analysis is the "hyperfine splitting" of the ground energy state of hydrogen atoms. As the verification and validation by Newton's final experiment, Newton's gravity constant *G* has been determined by this new approach, $G = 6.67221937(40) \times 10^{-11}$ N m$^2$ /kg$^2$. The result shows that, the gravitational constant *G* is really a universal constant in free space, and can be determined just by the more fundamental constants in physics.

## 1. Introduction

The problems of quantizing gravity as well as origin of gravitation are basic physics problems unsolved yet. Anyhow, the only characterization of gravitational strength is the Newtonian gravitation constant *G*, therefore, the value of *G* under the condition of no freely specifiable or adjustable parameters determined by any gravitational theory will be considered as the verification and validation by Newton's final experiment.

The official CODATA value for Newtonian constant of gravitation *G* in 2010 was given as $G = 6.67384(80) \times 10^{-11}$ N m$^2$ /kg$^2$,[1] in 2006, it was $G = 6.67428(67)\times 10^{-11}$ N m$^2$/kg$^2$.[2] However the value of *G* has been called into question frequently in recent years by new measurement from respected research. In 2000, Jens Gundlach and Stephen Merkowitz at the University of Washington in Seattle got the result[3] $G = 6.674215(14)\times 10^{-11}$ N m$^2$/kg$^2$; in 2009, researchers[4] led by Jun Luo of Huazhong University of Science and Technology in Wuhan, China, measured *G* a value of $6.67349(26)\times 10^{-11}$ N m$^2$/kg$^2$; back in 2005, one of the recommended values by the same group is $G = 6.67228\ (87)\times 10^{-11}$ N m$^2$ /kg$^2$; in 2010, Harold Parks and Faller at Sandia National Laboratories in Albuquerque, New Mexico, obtained a result[5] $G = 6.67234(21)\times 10^{-11}$ N m$^2$/kg$^2$. These results would only get further confusion and controversy [6], which strongly suggests that we have not completely understood the nature of gravity.

Furthermore, for the Newtonian gravity, we know how large object will act. Could the same laws keep success in atomic scales? In this study, a new semi-classical theory of quantum electrodynamics gravity (or simply "QED gravity" for short) and mechanism for origin of gravitational mass is developed. As a confirmation of this QED gravity theory, the Newtonian constant of gravitation *G* has been derived from the theory, that is $G = 6.67221937(40) \times 10^{-11}$ N m$^2$ /kg$^2$. This result would explain the inconsistency of different value of *G* measured above.

## 2. The quantization of electrostatic field

As yet, the quantization of electrostatic field is far from satisfied degree, but it is the foundation of the so-called unified field theory for the gravitation and the electromagnetism. According to the classical QED principle of the electrostatic field, the Coulomb's force between two electric charges comes from the exchange of the virtual photons. Electrostatic fields may be filled with such virtual photons, but these virtual photons are not able to be observed.

Unlike the classical QED field, we suggest that the field particles are real photons rather than virtual photons. However, in a quantum field theory of independent of the background, photons as the field particles cannot be defined on the coordinate space and time. In our problem, we are concerned with the force field, hence the quantized electromagnetic wave field should be defined in momentum space. Where, the photon is a wave train, the net momentum of the wave train is p = $\kappa\hbar$, here, ħ is Planck constant h divided by 2π; and ĸ represents the wave number.

If we consider a dimensionless momentum space, in such momentum space the quantized electromagnetic wave field would be described as the phase factor field. The momentum p = ĸ ħ is carried by wave train, ħ is carried by every period and represents the elementary unit of the wave train, just like the thread pitch of a spring. Thus, ĸ = $\omega/c$ = 2π / ħ , here, $\omega$ is angular frequency of wave train and c is the speed of light. Also, we define the Planck phase for one ħ is 1/ ĸ = ħ / 2π, although ħ is dimensionless here, it still has its numerical value in IS system.

Thus, a quantized phase factor can be defined as

$$e^{i\varphi(t,\kappa)} = e^{i(\omega t - \kappa \cdot \hbar/2\pi)} \tag{1}$$

where

$$\left|e^{i\varphi(t,\kappa)}\right| = \left|e^{i(\omega t - \kappa \cdot \hbar/2\pi)}\right| = 1 \tag{2}$$

The quantum field composed of these wave trains has the properties of electromagnetic wave, and it can be expressed in the form of Maxwell's wave

$$\vec{\boldsymbol{E}} = \vec{\boldsymbol{E}}_{\boldsymbol{0}}\, expi(\omega t - \kappa \cdot \hbar/2\pi) \tag{3}$$

After these wave trains are radiated from electric charge, the flux should obey Gauss' law

$$\oint\!\!\oint \vec{\boldsymbol{E}} \cdot \overrightarrow{\boldsymbol{ds}} = \oint\!\!\oint \vec{\boldsymbol{E}}_{\boldsymbol{0}}\, expi(\omega t - \kappa \cdot \hbar/2\pi) \cdot \overrightarrow{\boldsymbol{ds}} = q/\varepsilon_0 \tag{4}$$

The electric flux is the vertical component of electric field $\vec{\boldsymbol{E}}$ through the Gauss surface. According to Gauss' law, we have

$$\oint\!\!\oint \vec{\boldsymbol{E}} \cdot \overrightarrow{\boldsymbol{ds}} = 4\pi r^2 E_r = 4\pi r^2 E_0 expi(\omega t - \kappa \cdot \hbar/2\pi) = q/\varepsilon_0 \tag{5}$$

where $q$ is electron charge, and $\varepsilon_0$ is the permittivity of vacuum. By equations (5), we get

$$E_0 = \frac{q}{4\pi\varepsilon_0 r^2} expi[-(\omega t - \kappa \cdot \hbar/2\pi)] \tag{6}$$

Also by equations (5) and for

$$expi(\omega t - \kappa \cdot \hbar/2\pi) \cdot expi[-(\omega t - \kappa \cdot \hbar/2\pi)] = 1 \tag{7}$$

we obtain the Coulomb's law

$$E_r = q/4\pi\varepsilon_0 r^2 \tag{8}$$

However the wave field (3) gives the electrostatic field a form of propagation, in which the phase factor is *expi* ($\omega t$ − ĸ ħ / 2π) and propagates with the speed of light, but the formula (8) shows a form of electrostatic field. This is just the effect of the theory of radiation and absorption by Wheeler and Feynman, which suggested that the electric interaction between charges is through the advanced and the retarded waves [7]. That is that the advanced and retarded wave trains constitute the stand wave field in space, which acts as if there is an electrostatic field.

## 3. Interaction of two magnetic moments and the propagating electrostatic field

We defined the electrostatic field with a form of propagation. However, in classic electromagnetic theory there is no propagating electrostatic field. What is the quantum mechanics of propagating electrostatic field in QED? We choose hydrogen atoms as our model to practice such a quantum field theory. Hydrogen atom is the simplest atom in all of existence, and is usually in the lowest energy. In the ground state, the spin-spin interaction between electrons and protons can tear the ground state, which is called the hyperfine splitting. The quantum jumping between two states produces and radiates microwave photons with the angular frequency of ω, which had been studied perfectly with quantum mechanics. The hydrogen atom could emit or absorb the photon, which is a microwave photon rather than an optical one. By shining microwaves on hydrogen gas, we will find the absorption of energy at the angular frequency $\omega$, the frequency of microwave photons has been measured experimentally, the accurate result is $f = \omega / 2\pi = 1.420405751768(1)\times 10^{9}$ Hz,[8] in astronomy, this is the signature 21-centimeter emission line of hydrogen sources.

In the inside of hydrogen atom, microwave photons reflect between the electron and the proton. If a proton has sent photons, as wave trains, the propagating field (3) would be

$$\vec{\mathbf{E}}_{+} = \vec{\boldsymbol{E}}_{\mathbf{0}}\, expi(\omega t - \text{κ} \cdot \hbar/2\pi) \tag{9}$$

After absorbing those photons, in order to ensure the conservation of the angular momentum, the electron reflects photons with field backward

$$\vec{\mathbf{E}}_{-} = -\vec{\boldsymbol{E}}_{\mathbf{0}}\, expi[-(\omega t - \text{κ} \cdot \hbar/2\pi)] \tag{10}$$

here we have the minus sign because the negative charge of the electron. The Coulomb's force between positive and negative charges in hydrogen comes from the exchange of such microwave photons with the frequency $\omega$.

## 4. Destructive interfere microwave photons and the quantized gravitational field

In the case of outside of the hydrogen atom, these two kinds of photons propagate parallel forward along radius direction, but their phase has a half-wave difference in head and the tail of the wave trains. In momentum space, a half wave length is represented by $\hbar / 4\pi$ , therefore for the wave trains emitted from proton

$$\vec{\mathbf{E}}_{+} = \vec{\boldsymbol{E}}_{\mathbf{0}}\; expi[\omega t - (\text{κ} + 1/2) \cdot \hbar/2\pi + \hbar/4\pi] \tag{11}$$

and for the wave trains emitted from electron

$$\vec{\mathbf{E}}_{-} = -\vec{\boldsymbol{E}}_{\mathbf{0}}\; expi[\omega t - (\text{κ} + 1/2) \cdot \hbar/2\pi + \hbar/4\pi] \tag{12}$$

In the direction of connecting the electron and the proton, outside of the electron orbit, $\vec{\mathrm{E}}_{+}$ has a half wave length ahead, and $\vec{\mathrm{E}}_{-}$ has a half wave length delayed. These two simple harmonic wave trains propagate in the same direction. The total field after superimposition would be

$$\begin{aligned}\vec{\boldsymbol{E}} = \vec{\boldsymbol{E}}_{+} + \vec{\boldsymbol{E}}_{-} &= \vec{\boldsymbol{E}}_{\mathbf{0}}\; expi[\omega t - (\text{κ} + 1/2) \cdot \hbar/2\pi + \hbar/4\pi] \\ &\quad - \vec{\boldsymbol{E}}_{\mathbf{0}}\; expi[\omega t - (\text{κ} + 1/2) \cdot \hbar/2\pi + \hbar/4\pi] \\ &= \vec{\boldsymbol{E}}_{\mathbf{0}}\,(i\hbar/4\pi - i\hbar/4\pi) expi[\omega t - (\text{κ} + 1/2) \cdot \hbar/2\pi] = 0 \end{aligned} \tag{13}$$

The above result is because of $\hbar / 4\pi \ll 1$, thus there is

$$expi[\omega t - (\text{κ} + 1/2) \cdot \hbar/2\pi + \hbar/4\pi] = (1+i\hbar/4\pi) expi[\omega t - (\text{κ} + 1/2) \cdot \hbar/2\pi] \tag{14}$$

Nonetheless, no charge is at rest in nature. For hydrogen atom, since the radius of proton orbit is much smaller than that of electron orbit, for the viewpoint of electron, the charge of proton seems like concentrating on its orbit center. Hence, according to the special theory of relativity, the phase difference of two wave trains in the formula (13) requires a relativity shift. For the rest frame of reference to electron, the proton is moving at the speed of $v$ instead. The total field after superimposition is

$$\begin{aligned}\vec{\boldsymbol{E}} = \vec{\boldsymbol{E}}_{+} + \vec{\boldsymbol{E}}_{-} &= \vec{\boldsymbol{E}}_{\mathbf{0}}\,(i\hbar/4\pi\sqrt{1 - v^2/c^2} - i\hbar/4\pi) expi[\omega t - (\text{κ} + 1/2) \cdot \hbar/2\pi] \\ &= \vec{\boldsymbol{E}}_{\mathbf{0}}\,(i\hbar/4\pi\sqrt{1 - \alpha^2} - i\hbar/4\pi) expi[\omega t - (\text{κ} + 1/2) \cdot \hbar/2\pi] \\ &= \mathrm{i}(\hbar\alpha^2/8\pi)\vec{\boldsymbol{E}}_{\mathbf{0}}\, expi[\omega t - (\text{κ} + 1/2) \cdot \hbar/2\pi] \end{aligned} \tag{15}$$

where, $v$ is the speed of electron at the ground state in hydrogen atom, the $\alpha$ is defined as the fine structure constant and $\alpha = v/c \approx 1/137$ . By equations (6) and (7), the strength of field after superimposition is

$$E_r = (\hbar\alpha^2/8\pi)\,(q/4\pi\varepsilon_0 r^2) \tag{16}$$

Note that, here $\hbar$ is dimensionless by definition of such a dimensionless momentum space, it just has its numerical value in IS system.

Equation (16) is the formula of QED gravitational field, since the value of $\hbar$ is very small, the strength of field is extremely weak. This result makes atom electrically neutral and the near-perfect cancellation of electrical effect.

However, this leaked electromagnetic wave is just the carrier wave of gravity; it constitutes the gravitational interaction mechanism.

## 5. The QED gravity and the Newtonian gravitation constant *G*

In our question, to a hydrogen atom located in such field as (15) and (16), for the rest frame of reference to electron, there is a Lorentz force on electron by the interaction with field (15) and (16). When the force produced from exchanging photons there is a recoil effect, so we get twice the force

$$\mathrm{F_e} = -2qE_r = -(q^2/4\pi\varepsilon_0)(\hbar\alpha^2/4\pi r^2) \tag{17}$$

The minus sign is because the negative charge of the electron. We let this force equal to the Newton's gravitational force $\mathrm{F}_G$ between proton and electron

$$\mathrm{F_e} = -(q^2/4\pi\varepsilon_0)(\hbar\alpha^2/4\pi r^2) = F_G = -Gm_pm_e/r^2 \tag{18}$$

where $m_p$ is the rest mass of proton; $m_e$ is the rest mass of electron; $r$ is the distance between proton and electron, and *G* is the gravitational constant. The corresponding gravitational constant *G* is

$$G = (q^2/4\pi\varepsilon_0)(\hbar\alpha^2/4\pi\, m_pm_e) \tag{19}$$

But this is not the correct value of *G* in frame of reference to laboratory system. It is necessary to consider the relativistic and Doppler effect in radiation and absorption for the transformation of the reference frame from electron to laboratory system. For the stationary observer in laboratory, there is time dilation in a moving reference frame of electron at Bohr orbit, this is equivalent to slow down the speed of light in a moving reference frame of electron relative to the stationary observer in laboratory.

In addition, we have to consider the relativity frequency shift by Doppler effect. By shining microwaves on hydrogen gas, the angular frequency of microwave photons we measured is $\omega$. Anyhow that is the angular frequency which we observed or received in the frame of reference to laboratory system, the natural angular frequency of photons emitted from electron should be $\omega_0$ in the viewpoint of viewer in laboratory system. This is because the aberration effect of relativity movement between electron and proton. The aberration effect of constant circling movement produces a duration longitudinal Doppler effect, which is the same equivalence as the electron is always moving toward the proton with speed $v = \alpha c$. Therefore the proton will see photon of a higher frequency, the result is

$$\omega = \omega_0 \frac{1+v/c}{\sqrt{1-(v/c)^2}} = \omega_0 \sqrt{\frac{1+v/c}{1-v/c}} = \omega_0 \sqrt{\frac{1+\alpha}{1-\alpha}} \tag{20}$$

where, $v$ is the speed of electron at the ground state in hydrogen atom, $\alpha$ is the fine structure constant and $\alpha = v/c$. Letting

$$\beta = \sqrt{\frac{1+\alpha}{1-\alpha}} \tag{21}$$

we have

$$\omega_0 = \omega/\beta \tag{22}$$

If the speed of light is $c$ in free space of the frame of reference to laboratory system, the angular frequency of photons is $\omega$, and its wavelength is $\lambda$, according the definition, we have

$$c = \lambda\omega/2\pi \tag{23}$$

But to the frame of reference of moving electron, in the viewpoint of viewer in laboratory system, the natural angular frequency of photons emitted from electron would be reduced to $\omega_0$, there must have a lower apparent speed of light

$$c_0 = \lambda\omega_0/2\pi = \lambda\omega/2\pi\beta \tag{24}$$

By the definition of $c_0 = 1/\sqrt{\varepsilon\mu}$, and $c = 1 / \sqrt{\varepsilon_0\mu_0}$, here, $\varepsilon$ is the apparent permittivity reference to moving electron in the viewpoint of viewer of the reference frame to laboratory system, $\mu$ is the permeability accordingly, and $\mu = \mu_0$ in space of vacuum. Combine (23) and (24), we get

$$(c/ c_0)^2 = \varepsilon/\varepsilon_0 = \beta^2 \tag{25}$$

That is

$$\varepsilon = \varepsilon_0\beta^2 \tag{26}$$

Therefore, the relativity effect in radiation and absorption can be considered as the apparent changeable speed of light, although the angular frequency and the speed of light in the viewpoint of viewer of the static coordinate to electron is completely the same as the $\omega$ and $c$ rather than the $\omega_0$ and the $c_0$. This is the permanent principle of light velocity. Thus we get the correct QED gravity between proton and electron in frame of reference to laboratory system (the frame of reference in which the proton is at rest, because $m_p \gg m_e$).

$$F_G = \frac{F_e}{\beta^2} = -\left(\frac{q^2}{4\pi\varepsilon}\right)\frac{\hbar\alpha^2}{4\pi r^2}$$

$$= -(q^2/4\pi\varepsilon_0\beta^2)(\hbar\alpha^2/4\pi r^2 ) \tag{27}$$

The corresponding gravitational constant *G* is

$$G = (q^2/4\pi\varepsilon_0\beta^2)\,(\hbar\alpha^2/4\pi\ m_p m_\mu) \tag{28}$$

where, $m_\mu$ is the reduced mass of hydrogen atom

$$m_\mu = \frac{m_p m_e}{m_p + m_e}$$

Substituting numerical values into formula (28), where, that is
$\hbar = 1.054571726(47) \times 10^{-34}$; $q = 1.602176565(35) \times 10^{-19}$C; $\alpha = 7.2973525698(24) \times 10^{-3}$;
$m_e = 9.109\,382\,91(40) \times 10^{-31} kg$; $m_p = 1.672\,621\,777(74) \times 10^{-27} kg$; $\varepsilon_0 = 8.854187817 \times 10^{-12}$F $m^{-1}$.
We get the value of gravitational constant *G* in free space

$$G = 6.67221937(40) \times 10^{-11}\ \mathrm{N \cdot m^2/kg^2}$$

The result of formula (28) shows that the gravitational constant *G* is determined only by other fundamental physical quantity, and is not the same value everywhere; only in empty space (in free space) the *G* is a universal constant. This is because of the velocity of light is constant only in free space. However, there are no free space in the Earth's environment, so there is no a universal constant value of *G* in the Earth at all. This is why there are so many different value of *G* measured so far in different times and different places.

## 6. Conclusion

From above result we realize that the gravity force is the electromagnetic force and the mass is just the effect of quantum electrodynamics. The result shows that the mechanical piece of mass is not there at all, the mass is all electromagnetic in origin. Further, the fermions can now have mass, by interacting with the microwave photons associated microwave background radiation field, all the origins of the mass can be explained at low energy level, it is the ordinary phenomena of low energy, in the regions of zero point energy.

By the QED gravity theory, we get the accurate result of gravitational constant *G* in vacuum. Certainly the method of quantum resonance has the potential to make the value of *G* become one of the most remarkable accurate constants in physics. Although the result of QED gravity is derived in the case of "hyperfine splitting" of the ground energy state of hydrogen atoms, the value of gravitational constant *G* is independent of the angular frequency *ω* of the spin-spin interaction between electrons and protons and the concrete model of atoms, no matter with any substance and constitution form of objects, and no matter the interaction distance between objects even in atomic scale.